\documentclass[pdflatex,sn-mathphys-num]{sn-jnl}

\usepackage{graphicx}%
\usepackage{multirow}%
\usepackage{amsmath,amssymb,amsfonts}%
\usepackage{amsthm}%
\usepackage{mathrsfs}%
\usepackage[title]{appendix}%
\usepackage{xcolor}%
\usepackage{textcomp}%
\usepackage{manyfoot}%
\usepackage{algorithm}%
\usepackage{algorithmicx}%
\usepackage{algpseudocode}%
\usepackage{listings}%
\usepackage[utf8]{inputenc} 
\usepackage[T1]{fontenc}    
\usepackage{hyperref}       
\usepackage{booktabs}       
\usepackage{svg}
\usepackage{amsfonts}       
\usepackage{nicefrac}       
\usepackage{microtype}      
\usepackage{lipsum}
\usepackage{graphicx}
\usepackage{tikz}
\usepackage[T1]{fontenc}
\usepackage{enumitem}
\usepackage{forest, cuted}
\usepackage{xcolor}
\usepackage{multirow, makecell}
\usepackage{tcolorbox}
\usepackage{amsmath, amssymb}
\usepackage{url}

\usepackage{tabularx}
\usepackage{makecell}
\usepackage{xcolor}
\usepackage{pifont}

\usepackage{booktabs}
\usepackage{threeparttable}
\usepackage{graphicx}
\usepackage{array}
\usepackage{ragged2e}
\usepackage{subcaption}

\usepackage{threeparttable}
\usepackage{adjustbox}   
\usepackage[dvipsnames]{xcolor} 
\usepackage{caption} 

\definecolor{rec_color}{RGB}{200,36,35}
\definecolor{leaf_color}{RGB}{233,223,201}

\theoremstyle{thmstyleone}%
\theoremstyle{thmstyletwo}%

\theoremstyle{thmstylethree}%

\begin{document}

\title[Article Title]{Histopathology-centered Computational Evolution of Spatial Omics: Integration, Mapping, and Foundation Models}







\author[1,a]{\fnm{Ninghui} \sur{Hao}}

\author[1,a]{\fnm{Xinxing} \sur{Yang}}

\author[2]{\fnm{Boshen} \sur{Yan}}

\author[3]{\fnm{Dong} \sur{Li}}

\author[4]{\fnm{Junzhou} \sur{Huang}}

\author[5]{\fnm{Xintao} \sur{Wu}}

\author[6]{\fnm{Emily S.} \sur{Ruiz}}

\author[7]{\fnm{Arlene} \sur{Ruiz de Luzuriaga}}

\author[3,b]{\fnm{Chen} \sur{Zhao}}

\author*[1,b]{\fnm{Guihong} \sur{Wan}}\email{gwan@uchicago.edu}

\affil[1]{\orgdiv{Institute for Population and Precision Health}, \orgname{University of Chicago}, \orgaddress{
\city{Chicago}, 
\state{IL}, \country{USA}}}


\affil[2]{\orgdiv{Department of Computational Biology}, \orgname{Carnegie Mellon University}, \orgaddress{
\city{Pittsburgh}, 
\state{PA}, \country{USA}}}

\affil[3]{\orgdiv{Department of Computer Science}, \orgname{Baylor University}, \orgaddress{
\city{Waco}, 
\state{TX}, \country{USA}}}

\affil[4]{\orgdiv{Department of Computer Science and Engineering}, \orgname{The University of Texas at Arlington}, \orgaddress{
\city{Arlington}, 
\state{TX}, \country{USA}}}

\affil[5]{\orgdiv{Department of Electrical Engineering and Computer Science}, \orgname{University of Arkansas}, \orgaddress{
\city{Fayetteville}, 
\state{AR}, \country{USA}}}

\affil[6]{\orgdiv{Department of Dermatology}, \orgname{Brigham and Women's Hospital, Harvard Medical School}, \orgaddress{
\city{Boston},  
\state{MA}, \country{USA}}}

\affil[7]{\orgdiv{Section of Dermatology}, \orgname{University of Chicago}, \orgaddress{
\city{Chicago}, 
\state{IL}, \country{USA}}}

\affil[a]{Designated co-first authors}
\affil[b]{Designated co-senior authors}



\abstract{
Spatial omics (SO) technologies enable spatially resolved molecular profiling, while hematoxylin and eosin (H\&E) imaging remains the gold standard for morphological assessment in clinical pathology. Recent computational advances increasingly place H\&E images at the center of SO analysis, bridging morphology with transcriptomic, proteomic, and other spatial molecular modalities, and pushing resolution toward the single-cell level.
In this survey, we systematically review the computational evolution of SO from a histopathology-centered perspective and organize existing methods into three paradigms: integration, which jointly models paired multimodal data; mapping, which infers molecular profiles from H\&E images; and foundation models, which learn generalizable representations from large-scale spatial datasets. We analyze how the role of H\&E images evolves across these paradigms from spatial context to predictive anchor and ultimately to representation backbone in response to practical constraints such as limited paired data and increasing resolution demands.
We further summarize actionable modeling directions enabled by current architectures and delineate persistent gaps driven by data, biology, and technology that are unlikely to be resolved by model design alone. Together, this survey provides a histopathology-centered roadmap for developing and applying computational frameworks in spatial omics.
}


\maketitle

\section{Introduction}\label{sec1}

Spatially resolved transcriptomics (ST) was named Method of the Year 2020~\cite{marx2021method}, followed by spatial proteomics (SP) in 2024~\cite{karimi2024method}. Together, these spatial omics (SO) technologies extend bulk and single-cell profiling into a spatially organized setting, enabling comprehensive analysis of tissue architectures, cellular interactions, and microenvironmental features in development and disease. 
In typical workflows, SO technologies 
provide (i) molecular measurements (e.g., gene expression or protein abundance matrices) and (ii) spatial coordinates of measured regions, often accompanied by matched hematoxylin and eosin (H\&E) whole-slide images (WSIs) that provide a common morphological backdrop for 
interpretation and alignment. 
In this sense, SO complements H\&E images by adding spatially resolved molecular profiles on top of a long-established morphological readout.

H\&E-stained histological images remain the gold standard in clinical pathology, providing visualization of tissue architecture and cellular morphology~\cite{fischer2008hematoxylin}. 
Beyond its routine clinical use, H\&E images preserves rich context from subcellular to tissue-level organization and is readily available across platforms, cohorts, and disease contexts.
In this context, H\&E images increasingly function as a shared morphological anchor for SO to connect heterogeneous spatial measurements across different platforms, technologies, and studies.
Importantly, the role of H\&E images in SO computation evolves. Early computational approaches prioritized molecular matrices and treated H\&E images primarily as a background that supplies spatial coordinates or rough tissue masks~\cite{yang2023revealing}. 
More recent work elevated H\&E images to a first-class modality, using them for representation learning, cross-modal alignment, and even pretraining objectives that explicitly model morphology-omics correspondence~\cite{li2025multi}.
Taken together, these developments suggest that a central axis of progress in SO computation is how methods interpret and operationalize morphological information from H\&E images, often more so than the choice of backbone architecture alone.

\begin{figure}[H]
    \centering
    \includegraphics[width=\textwidth, trim=0cm 0cm 0cm 0, clip]{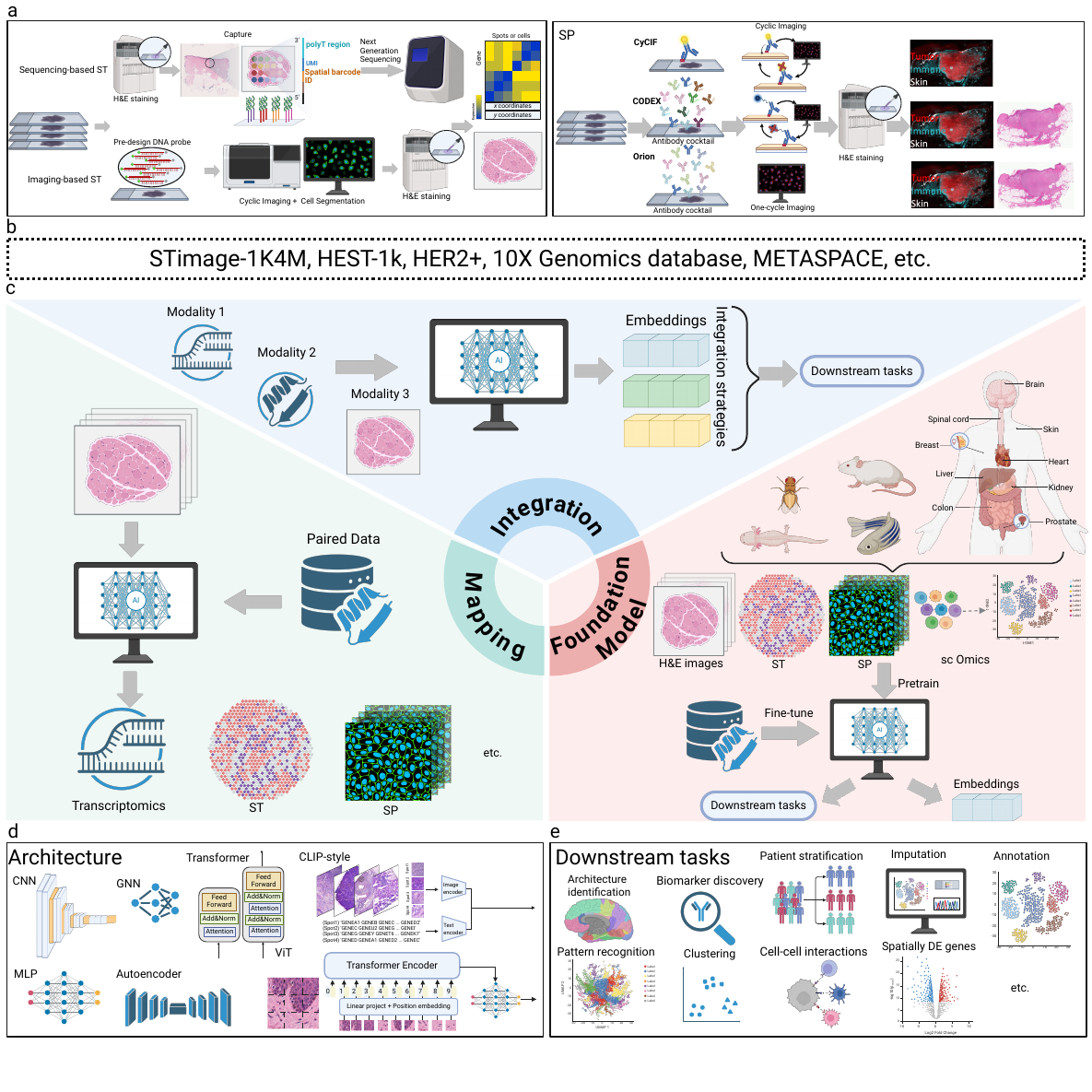}
    \caption{Computational landscape of spatial omics and H\&E image analysis.\\
\textbf{a.} Illustration of general ST and SP techniques.\\
\textbf{b.} Commonly used public SO datasets. 
Details are in the Additional file 1: Table S1.\\
\textbf{c.} Three categories of current methods: Integration methods align and fusion multiple data modalities; 
Mapping methods focus on H\&E-to-X prediction or imputation; 
Foundation models are pretrained on large heterogeneous datasets and fine-tuned for spatial omics tasks.
Additional file 1: Table S2 presents details of all reviewed methods.\\
\textbf{d.} Example architectures used in integration, mapping, and foundation models.\\
\textbf{e.} Representative downstream tasks.\\
Abbreviations: 
SO, spatial omics; 
ST, spatial transcriptomics; SP, spatial proteomics; sc, single-cell; 
CNN, convolutional neural network; 
MLP, multilayer perceptron; GNN, graph neural network; GCN, graph convolutional network; 
ViT, vision transformer; CLIP, contrastive language-image pretraining; 
DE, differentially expressed.
}
\label{fig:introF1}
\end{figure}

In this survey, we categorize computational methods for H\&E 
and SO analysis into three classes based on the interaction of different modalities and the development evolution (Fig.~\ref{fig:introF1}).
First, \textbf{integration} methods combine multimodal SO measurements (with H\&E images), where H\&E typically provides morphological context and spatial coordinates. 
Second, \textbf{mapping} methods emphasize cross-modal inference, particularly H\&E-to-X prediction, where H\&E is treated as the primary observable and X denotes molecular modalities (ST, SP, and beyond) that may be expensive, noisy, or missing. 
Third, emerging \textbf{foundation models} aim to learn general-purpose representations from large-scale spatial datasets, which operationalize H\&E usage at scale by pretraining on large collections of paired or weakly paired image-omics data, with the goal of applying representations to diverse downstream tasks.

\begin{figure}[b]
    \centering
    \includegraphics[width=\textwidth, trim=0cm 0cm 0cm 0, clip]{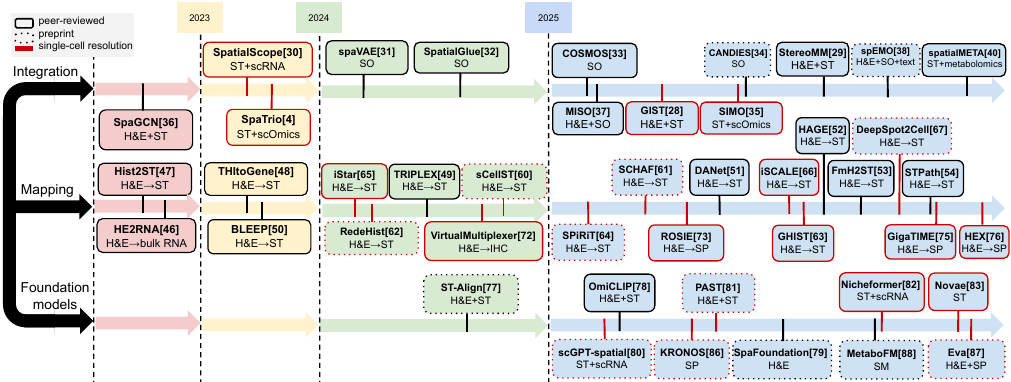}
    \caption{A roadmap of computational methods of spatial omics and H\&E image analysis: Integration, Mapping, and Foundation Models.
Solid boxes denote peer-reviewed publications; dashed boxes indicate preprints. Red boxes denote methods with explicit single-cell-level support (These methods are annotated as `Yes' in the `Single-cell level' column in the Additional file 1: Table S2). 
Some methods that do not explicitly incorporate H\&E images (e.g., SpaTrio~\cite{yang2023revealing}) are presented for completeness.
Abbreviations: 
SO, spatial omics; 
ST, spatial transcriptomics; 
SP, spatial proteomics; 
SM, spatial metabolomics; 
scOmics, single-cell omics; 
scRNA, single-cell RNA; 
IHC, immunohistochemistry.
}
    \label{fig:roadmapF2}
\end{figure}

From the H\&E perspective, these three paradigms represent a natural progression. 
When paired multimodal data are available, integration utilizes 
H\&E morphology to enhance spatial molecular analysis; 
when modalities are missing or costly, mapping uses H\&E images to infer 
missing molecular profiles; 
and with sufficient data accumulation, 
foundation models are trained to learn representations that support diverse downstream tasks 
with improved 
robustness and 
generalization by leveraging ubiquitous, golden-standard H\&E images, informed by spatial SO data.
We also summarize a roadmap from both temporal and conceptual views of how the field is evolving toward H\&E-anchored, single-cell-aware multimodal analysis (Fig.~\ref{fig:roadmapF2}).

A major driver of this evolution is the push toward single-cell and subcellular resolution. 
Recent technical advances 
have transitioned SO from spot-level readouts to cellular and even subcellular measurements, including emerging platforms such as Visium~HD~\cite{oliveira2025high}. 
In earlier ST techniques such as Visium~\cite{staahl2016visualization}, each spot aggregates expression from multiple cells, typically without explicit cell-type annotation or true single-cell profiles, which can obscure key biological variation and complicate mechanistic interpretation. 
%
%
Single-cell (and subcellular) resolution has become a necessity; however, comprehensive single-cell SO 
datasets
remain constrained by cost, throughput, and experimental complexity. 
%
A growing share of the bottleneck shifts to computation: 
how to exploit heterogeneous modalities to 
recover cell-level characteristics, 
link morphologies to molecular programs, and translate spatial readouts into biological insights and 
clinical
applications~\cite{williams2022introduction}. In this context, fully leveraging cellular-level morphology from H\&E images is a particularly promising strategy for enabling robust, generalizable single-cell representations through H\&E-centered multimodal learning.

Although existing surveys have summarized SO literature, they often lack in-depth insights and comprehensive synthesis across integration, mapping, and spatial foundation models (Table~\ref{tab:survey_comparison}). 
Prior surveys have largely focused on integration strategies of modality embeddings~\cite{luo2025deep}, on specific tasks (e.g., deconvolution~\cite{gaspard2025cell}), or on clinical applications~\cite{lee2025spatial}. By contrast, our survey treats H\&E images as an anchor and uses them to connect methodological threads across paradigms and across spatial resolutions.
The key contributions of our survey are summarized as follows:

\begin{itemize}[noitemsep,topsep=5pt]
    \item \textbf{Reframing the computational evolution of H\&E-centered SO analysis.}
We provide an H\&E-driven perspective to synthesize the computational evolution of SO from multimodal integration to H\&E-to-X mapping and, more recently, foundation models. We explicitly characterize how each paradigm leverages H\&E images, ranging from spatial/morphological context to conditional signal for cross-modal translation and finally to a representation anchor in large-scale pretraining, and how these shifts are shaped by practical constraints such as missing modalities and limited paired data.

    \item  \textbf{Current failure modes and future direction implication.} 
We systematically summarize common failure modes in current SO methods and distinguish challenges that can be mitigated through methodological and architectural design from fundamental limitations imposed by biological complexity, data availability, and technological constraints. This analysis clarifies the realistic scope of algorithmic improvements and informs future research directions beyond incremental model refinement.

    \item \textbf{Comprehensive and up-to-date coverage of spatial foundation models.} 
With the increasing amount of SO data, a variety of foundation models have been developed and pre-trained on large-scale paired or weakly paired samples. We provide an up-to-date overview of these models, detailing their training datasets and model architectures, and discuss how their learned representations can be transferred to downstream spatial analysis tasks and concrete research use cases.
\end{itemize}

\newcommand{\bluetick}{\textcolor{green}{\ding{51}}}     
\newcommand{\redx}{\textcolor{red}{\ding{55}}}         
\newcommand{\partialmark}{\textcolor{blue}{\ding{109}}} 

\begin{table}
\centering
\caption{Comparison of survey articles by year, 
resolution, foundation-model coverage, H\&E integration strategies, and clinical applications.}
\label{tab:survey_comparison}
 
\setlength{\tabcolsep}{1pt} 
\renewcommand{\arraystretch}{1.2}

\begin{tabular}{lcccccc}
 
\toprule
 
\textbf{Survey} &
\textbf{Year} &
\makecell[c]{\textbf{Resolution}} &
\makecell{\textbf{FM}\\\textbf{Coverage}} &
\makecell{\textbf{H\&E Integration}\\\textbf{Strategies}} &
\makecell{\textbf{Clinical}\\\textbf{Applications}} \\
 
\midrule
 
 
Zahedi et al. \cite{zahedi2024deep}
& 2024  & spot & \partialmark & \redx  & \redx \\
 
Dezem et al. \cite{dezem2024spatially}
& 2024 & single-cell & \partialmark & \redx  & \partialmark \\
 
Sun et al. \cite{sun2024comprehensive}
& 2024  & spot & \redx & \redx  & \redx \\
 
Liu et al. \cite{liu2024spatial}
& 2024  & spot & \partialmark & \redx  & \partialmark \\
 
Luo et al. \cite{luo2025deep}
& 2025  & spot & \partialmark & \redx  & \redx \\
 
Jana et al. \cite{jana2025bridging}
& 2025  & spot & \redx & \partialmark  & \bluetick \\
 
Lee et al. \cite{lee2025spatial}
& 2025  & spot & \partialmark & \partialmark  & \bluetick \\
 
Gaspard-Boulinc et al. \cite{gaspard2025cell}
& 2025  & spot \& single-cell & \redx & \redx  & \partialmark \\
 
\midrule\midrule
 
\textbf{This Survey}
& 2025  & spot \&single-cell & \bluetick & \bluetick  & \bluetick \\

\bottomrule
\end{tabular}
 
\vspace{1ex}
{\footnotesize \raggedright \textbf{Note:} Green \bluetick{} denotes ``yes'', red \redx{} denotes ``no''; blue \partialmark{} denotes partial coverage. Resolution includes single-cell and spot levels. FM: foundation model.\par}
\end{table}

\section{Spatial omics technologies and datasets}\label{sec:tech}

\subsection{Technology background: resolution and data quality}
ST is arguably the most widely used class of SO technologies and can be broadly categorized into sequencing-based ST and imaging-based ST (Fig.~\ref{fig:introF1}a). 
Both approaches generate spatially resolved gene expression data 
alongside corresponding H\&E images. 
In this subsection, we focus on how these two technology types differ in spatial resolution, data quality, and practical limitations.

Sequencing-based ST captures transcriptome-wide profiles (around 15-30k genes) using unique barcodes that correspond to predefined regions (`spots') on tissue slides, followed by next-generation sequencing. 
Spatial resolution is therefore determined by the relationship between spot size and individual cells. 
In practice, each ST spot typically contains multiple cells, and the resulting expression profile represents a mixture of transcripts from those cells (e.g., Visium~\cite{staahl2016visualization}, which uses a 55~\(\mu\)m spot diameter with 100~\(\mu\)m center-to-center spacing). 
Although advanced techniques employ spot sizes comparable to or smaller than a single cell to approach cellular or subcellular resolution 
(e.g., Visium HD~\cite{oliveira2025high} with 2~\(\mu\)m \(\times\) 2~\(\mu\)m continuous barcoded spots), 
the resulting measurements may still be ``pseudo'' single-cell expression because each spot can capture RNA from partial cells, adjacent cells, and/or intercellular gaps.  
Another limitation is the inherently small tissue capture area (Visium, 6.5 mm × 6.5 mm or 11 mm × 11 mm), restricting spatial measurements to selected fields of view 
(FOVs) 
rather than the entire tissue section. 
In sequencing-based ST workflows, H\&E staining is typically performed before RNA capture, preserving diagnostic-grade H\&E images.
Overall, sequencing-based ST provides (near) whole-transcriptome measurements and high-quality H\&E images, while spatial resolution depends on the specific platform.

Imaging-based ST techniques, such as MERSCOPE~\cite{moffitt2018molecular}, CosMx~\cite{he2022high} and Xenium~\cite{janesick2023high}, provide cellular resolution and can handle moderately larger tissues with a compensation of a limited number of genes and longer image scanning time. After tissue preparation, designed probes targeting specific genes are hybridized to the section, followed by cyclic imaging to capture RNA signals. At the beginning of the imaging workflow, DAPI-stained images are typically acquired to localize DNA (i.e., nuclei), 
providing a reference for subsequent cell segmentation. 
H\&E staining is performed after cyclic imaging of all probes.
Image-based ST can achieve true single-cell resolution by leveraging cell segmentation algorithms and sometimes reach subcellular resolution 
(contingent on accurate cell segmentation). 
However, the number of measured genes is constrained by probe design, which is limited compared to the (near) transcriptome-wide coverage in sequencing-based ST. In addition, the quality of H\&E images is affected by the cyclic staining and imaging steps. Current algorithms often treat H\&E images from both techniques as the same, which could be considered and improved in future work. 


For downstream algorithm development and analysis, H\&E images are represented at multiple scales using distinct terminology.
In general, tissue blocks are sectioned into thin (e.g., 5 $\mu$m) sections and mounted on glass slides. 
After H\&E staining, these are referred to as H\&E-stained slides.
H\&E images typically denote digitized WSIs scanned from H\&E-stained slides, which are often several gigabytes in size per image. 
For computational modeling, H\&E WSIs are commonly subdivided into smaller images using a sliding-window strategy, referred to patches or tiles.
At the finest spatial scale, an H\&E pixel represents the raw intensity value at a single spatial location.

SP also encompasses multiple technology classes, including imaging-based SP (e.g., multiplex immunofluorescence, mIF) and mass spectrometry (MS)-based SP (e.g., imaging mass cytometry, IMC). 
While immunohistochemistry (IHC) remains the gold standard for assessing protein expression in clinical settings, it offers limited multiplexing. Modern SP techniques have evolved from traditional mIF to 
CyCIF~\cite{lin2018highly} and CODEX~\cite{van2022multiplex}, 
and more recently, toward high-throughput single-pass systems like Orion~\cite{lin2023high}. 
IMC~\cite{chang2017imaging} utilizes metal-tagged antibodies coupled with laser ablation and mass cytometry readout, thereby circumventing spectral overlap inherent to fluorescence-based imaging. In contrast, Deep Visual Proteomics (DVP)~\cite{mund2022deep} integrates high-resolution imaging, laser microdissection, and ultra-sensitive MS analysis to enable antibody-independent greater proteome coverage. Next, we briefly review several imaging-based SP techniques.

Cyclic immunofluorescence (CyCIF)~\cite{lin2018highly} is an iterative imaging method that achieves high-plex protein profiling using conventional, fluorophore-conjugated antibodies. It operates through repeated cycles of antibody staining, fluorescence imaging, and signal inactivation via chemical bleaching or fluorophore quenching. This enables the detection of dozens of markers on a single tissue section. While CyCIF is highly accessible, the repeated chemical bleaching can compromise tissue integrity, making subsequent H\&E staining of the same section impractical. In practice, H\&E images are therefore typically acquired from adjacent tissue sections.

CODEX~\cite{van2022multiplex} 
from Akoya employs a DNA-barcoding strategy. Unlike CyCIF, CODEX processes a tissue section stained with a multiplexed antibody cocktail (e.g., 40-100 antibodies with barcodes), followed by cyclic imaging where fluorescently labeled complementary DNA probes are iteratively hybridized and removed. DAPI-stained images are acquired at the cyclic imaging stage to visualize nuclear morphology and support subsequent cell segmentation. 
Similar to imaging-based ST, H\&E staining is performed after cyclic imaging, leading to degraded H\&E image quality. In summary, CODEX takes one tissue section and outputs multiplexed protein expression maps with a cell segmentation mask and a suboptimal H\&E image.

Orion~\cite{lin2023high} from RareCyte is a recently introduced single-pass imaging platform designed to better preserve tissue morphology.
Tissue sections are incubated with a cocktail of up to 20 antibodies, each conjugated to a specific fluorophore. Single-pass mIF imaging then captures all protein signals simultaneously, followed by fluorescence bleaching to enable subsequent H\&E staining. 
In this workflow, a less degraded H\&E image can be obtained from the same section. Orion therefore provides up to 20 spatial protein measurements and high-quality H\&E images from the same section, whereas CODEX supports higher multiplexing but yields more degraded H\&E images.

Together, sequencing-based ST provides near-whole-transcriptome measurements at spot-level resolution and high-quality H\&E images, whereas imaging-based ST achieves true single-cell resolution through cell segmentation but is limited to targeted genes and produces degraded H\&E images. 
SP exhibits a similar trade-off, with cyclic imaging methods such as CODEX offering high multiplexing at the cost of H\&E image quality, 
while single-pass techniques such as Orion preserve tissue morphology and H\&E quality with reduced multiplexing.

\subsection{Dataset summary}
We summarize the commonly
used public 
SO datasets 
by the reviewed computational methods in the Additional file 1: Table S1, organized by technique, modality, species, tissue sites, and access links. 

Most ST datasets could be found in the 10x Genomics and Gene Expression Omnibus (GEO) repositories. 
Canonical human ST datasets include the dorsolateral prefrontal cortex (DLPFC)~\citep{maynard2021transcriptome} and HER2-positive breast cancer (HER2ST)~\citep{andersson2021spatial}.
%
Broadly, ST studies utilize a range of organ tissues, including prostate, heart, colorectal, stomach, and pancreas across both healthy and diseased cohorts.
Mouse tissue benchmarks are equally prevalent across various brain regions, including the cortex, olfactory bulb, and anterior brain, as well as embryos and pups. 
Publicly available datasets for SP, spatial metabolomics (SM), and spatial epigenomics (SE) are also increasingly integrated into benchmarking pipelines. Together, these datasets serve as the dominant ``workhorse'' benchmarks, driving advancements in multimodal integration, H\&E-to-X mapping, and the pretraining of emerging spatial foundation models.

\begin{figure}[!t]
    \centering
    \includegraphics[width=\textwidth, trim=0cm 0.8cm 0cm 0cm, clip]{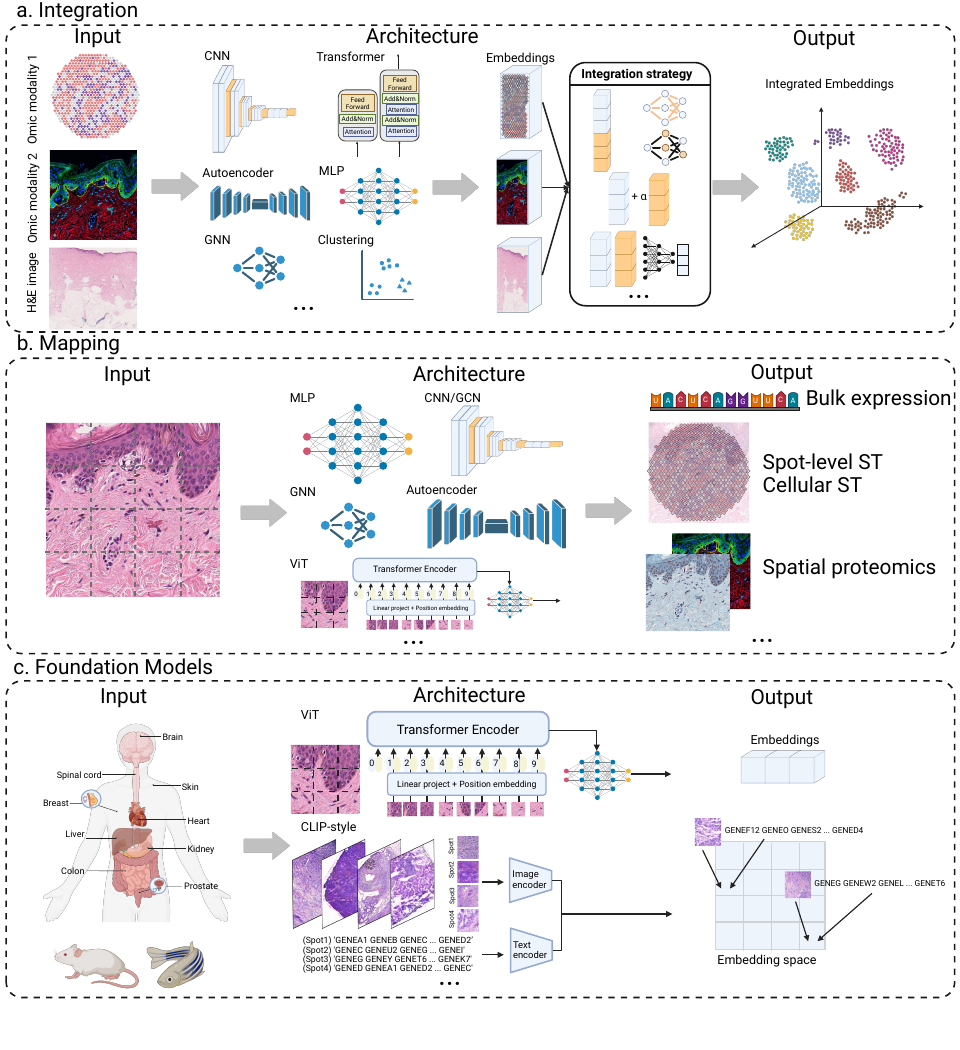}
    \caption{Integration, Mapping, and Foundation models. 
    Abbreviations: 
    CNN, convolutional neural network; 
    GNN, graph neural network; 
    MLP, multilayer perceptron;
    ViT, vision transformer;
    ST, spatial transcriptomics;
    CLIP, contrastive language-image pretraining.
  }
    \label{fig:methodF3}
\end{figure}

\section{Evolutionary trends of computational methods}

SO has rapidly evolved since the early 2020s, both in terms of experimental platforms and computational analysis. 
In the earliest stage, computational tools were primarily embedded within ST workflows to analyze sequencing outputs together with matched H\&E images. 
Later efforts extended to integrating ST data across platforms and with other SO modalities. We refer to this class of approaches as integration methods.
As datasets and applications continued to grow, new challenges emerged. 
Due to the scarcity of fully paired datasets and the desire to deconvolute spots to single-cell resolution, ``mapping'' methods were developed to infer information from one modality to another, especially from H\&E images to others. 
Integration and mapping methods have since co-evolved. We define integration as jointly modeling multiple modalities that are observed for the same samples, whereas mapping learns an explicit cross-modal function for information inference when paired measurements are incomplete or missing.
However, these methods are task-specific and lack generalizability. 
More recently, ideas from foundation models have been brought into SO analysis, enabled by the growing availability of large-scale datasets. Representations from foundation models can be reversely utilized by integration and mapping methods. 
Taken together, 
we organize the evolution of computational methods for SO into three categories: \textbf{integration, mapping, and foundation models} (Fig.~\ref{fig:methodF3}). 
We conclude that in this evolution, 
the role of H\&E images has progressively shifted
from serving as spatial backgrounds for coordinates, 
to acting as predictive anchors for cross-modal inference, 
and ultimately to functioning as a representation backbone that encodes morphological information for diverse downstream tasks.
%
For a comprehensive comparison of all reviewed methods, 
see the Additional file 1: Table S2.

\subsection{Integration methods}

Integrative methods aim to synthesize a comprehensive 
molecular profile from different modalities, revealing the intricate biological mechanisms from tissue architecture and molecular expression. 
A recent survey~\cite{luo2025deep} summarizes integration strategies.
The development of this field has shifted from geometry-only representations, which rely solely on spatial coordinates (e.g.,~\cite{wu2024discovery}), toward morphology-informed integration. Methods that explicitly encode H\&E patches to capture cellular morphology and local spatial interactions have generally achieved richer and more biologically relevant spatial representations (e.g.,~\cite{ge2025deep, luo2025stereomm}).

Early integration efforts primarily focused on cross-platform ST datasets or different SO modalities, where H\&E images served as a spatial reference or validation for cell boundary alignment. For instance, SpatialScope~\cite{wan2023integrating} introduces a 
framework that leverages paired single-cell RNA sequencing (scRNA-seq) as a reference to deconvolute spot-level ST into single cells. By combining nucleus segmentation on H\&E images, cell-type identification inferred from scRNA-seq, and gene expression decomposition, it achieves cellular resolution through a probabilistic deconvolution of spot compositions. Several similar methods 
build graphs of ST (or SO) data with specific cell-to-cell mapping algorithms to deconvolute (e.g.,~\cite{yang2023revealing
}). 
Subsequently, more sophisticated architectures were adopted to handle complex integrations, including 
variational autoencoders (VAE) (spaVAE~\cite{tian2024dependency}), 
attention mechanisms (SpatialGlue~\cite{long2024deciphering}), 
contrastive discriminators (COSMOS~\cite{zhou2025cooperative}),
and conditional diffusion models (CANDIES~\cite{liu2025cross}).
While SIMO~\cite{yang2025spatial} further pushed this boundary to integrate single-cell multi-omics, 
these methods largely utilized H\&E images as spatial structure referencing or result validation rather than a rich source of morphological information.

A conceptual shift emerged as methods began to incorporate H\&E patches into model architectures for morphological embeddings.
SpaGCN~\cite{hu2021spagcn} is an early example of this paradigm that introduces a graph convolutional framework to integrate ST with H\&E image patches for downstream analyses, particularly the identification of spatially variable genes.
GIST~\cite{ge2025deep} continued to leverage H\&E patches for spatial profiling through contrastive learning and knowledge distillation enhanced feature extraction, 
while StereoMM~\cite{luo2025stereomm} utilized graph autoencoders to align transcriptomic data with morphology and spatial locations.
A representative algorithm is MISO~\cite{coleman2025resolving}, designed to integrate H\&E images, ST, and multiple SO data. 
MISO constructs a 
graph adjacency matrix for each input modality and encodes 
individual modalities into embeddings via separate autoencoders. 
By encoding each modality 
and 
focusing on inter-modality interaction features, MISO preserves the unique contributions of each data type while filtering out low-quality modalities to prevent noise propagation. 
This trend has further evolved with spEMO~\cite{liu2025spemo}, which incorporates embeddings from Pathology Foundation Models (PFMs)~\cite{li2025multi, li2025survey}, Large Language Models (LLMs), and spatialMETA~\cite{tian2025integrating}, 
enabling the vertical and horizontal integration of H\&E images, ST, and SM across diverse samples.

Beyond methodological innovations, 
the proliferation of SO technologies requires standardized infrastructures and platforms (e.g., SpatialData~\cite{marconato2025spatialdata} and Thor~\cite{zhang2025thor}) to mitigate technical variation and cross-dataset reuse limitations. 
SpatialData~\cite{marconato2025spatialdata} addresses this by defining a unified storage format optimized for large-scale multimodal datasets. 
The SpatialData framework supports interactive analysis, including visualization, annotation, and landmark-based registration of datasets, and integrates seamlessly with the scverse ecosystem~\cite{virshup2023scverse}  
for downstream spatial analyses~\cite{wolf2018scanpy,wan2024spatialcells}. 

In summary, integration methods evolve 
from bulk to cellular resolution and 
from ST to multiple SO modalities, 
progressively advancing architectures to tackle batch effects, noise, and resolution limitations. 
However, these methods fundamentally depend on paired modalities, and H\&E often serves only as spatial context rather than a fully exploited morphological signal. 
As paired data become the limiting factor, 
mapping approaches emerge to extend these frameworks to infer missing modalities or refine resolution from partially observed data.

\subsection{Mapping methods}

Beyond paired integration, mapping methods relax the requirement of co-measured modalities by learning 
cross-modal inference functions. We name these approaches ``mapping'' 
because they establish a 
correspondence between modalities, 
translating signals across modalities. 
In this paradigm, H\&E images are commonly used as the anchor to infer molecular profiles from morphologies. Accordingly, we focus on H\&E-to-X mapping, where X can include (spatial) transcriptomics, proteomics, and other SO modalities.

Early mapping methods focused on transcriptomic prediction from H\&E images, spanning bulk expression (e.g., HE2RNA~\cite{schmauch2020deep} 
and 
spot-level ST 
(e.g.,~\cite{zeng2022spatial,
jia2023thitogene,
chung2024accurate,
xie2023spatially,
wu2025danet,dang2025hage,wang2025fmh2st,huang2025stpath}). 
Many of these approaches leverage 
convolutional neural networks (CNNs)~\cite{mondol2023hist2rna, chung2024accurate},
transformers~\cite{zeng2022spatial, jia2023thitogene},
and contrastive learning~\cite{xie2023spatially,wu2025danet}.
BLEEP~\cite{xie2023spatially} is the first to introduce a CLIP-based~\cite{radford2021learning} framework for H\&E-to-ST prediction. 
Despite architectural advances, predictive performance often remained modest, in part because many methods did not explicitly constrain the correspondence between paired modalities in their architectures.
More recently, generative and foundation model-based designs have been 
introduced~\cite{dang2025hage,
huang2025stpath}. 
STPath~\cite{huang2025stpath} is a generative model that bridges H\&E and ST through a PFM backbone~\cite{xu2024whole}, jointly encoding histological images, gene expression, organ type, and sequencing technology via a geometry-aware transformer to predict transcriptome-wide expression from H\&E images.
It was trained on a curated dataset of paired H\&E images and ST annotations from HEST-1K~\cite{jaume2024hest} and STImage-1k4m~\cite{chen2024stimage} 
with tailored noise schedules. 
Together, bulk- and spot-level mapping methods establish a foundation for morphology-driven transcriptomic inference, which 
naturally 
motivates subsequent extensions toward cell-level resolution.

Because individual cells are directly observable on H\&E images, subsequent work further refined mapping toward cellular resolution, enabling single-cell gene expression and cellular ST prediction~\cite{%
chadoutaud2024scellst,
comiter2024inference,
zhong2024spatial,
fu2025spatial,
zhao2024inferring
}. 
%
RedeHist~\cite{zhong2024spatial} leverages scRNA-seq references via a deconvolution algorithm and adopts a U-net framework to extract pixel-level features from H\&E images, mitigating the instability introduced by patch-based representations from irregular cell morphology. These components enable RedeHist to predict single-cell gene expression from H\&E images.
Similarly,
SPiRiT~\cite{zhao2024inferring} infers single-cell spatial gene expression from H\&E images, with an emphasis on interpretable and trustworthy prediction. It adopts a vision transformer (ViT) and introduces Training and Validation Attention Consistency (TAVAC), 
a
training metric that guides the model to select consistent high-attention regions.
By calculating TAVAC, SPiRiT produces robust predictions of disease-relevant and non-disease gene biomarkers, as well as tumor region annotations concordant with expert assessment. Nevertheless, while SPiRiT reports strong performance on its evaluated datasets, broader validation across tissue types, disease contexts, and imaging modalities remains necessary to establish generalization.

In parallel, related studies have focused on super-resolution (SR), which aim to computationally deconvolve or impute expression signals beyond the physical resolution limits of spatial sequencing techniques. 
iStar~\cite{zhang2024inferring} is a representative SR gene expression prediction model trained with paired H\&E and ST data using a pretrained hierarchical ViT architecture. Its key step is leveraging paired H\&E patches and ST sections to reconstruct unobserved, higher-resolution gene expression. Owing to strong performance in high-resolution tissue architecture annotation and SR gene expression prediction, iStar has become a common benchmark for subsequent SR approaches. 
iSCALE~\cite{schroeder2025scaling} uses scRNA-seq as a reference, combining a hierarchical ViT feature extractor and feedforward MLP to infer cellular ST from H\&E images. DeepSpot2Cell~\cite{nonchev2025deepspot2cell} adopts a DeepSet-based architecture and leverages pretrained pathology foundation models,
whose choice could be tailored to specific tasks, 
such as UNI~\cite{chen2024towards} or H-Optimus-0~\cite{saillardh}. Nevertheless, benchmarks suggest that SR models (e.g., iStar~\cite{zhang2024inferring}, scstGCN~\cite{xue2024inferring}) exhibit limited generalization to unseen samples beyond the training distribution.

Despite these advances, an open challenge in mapping is establishing reliable cross-modal correspondence. To address this, contrastive learning (CL) has been introduced to strengthen alignment by pulling matched (``positive'') pairs together while pushing unmatched (``negative'') pairs apart in a shared embedding space. DANet~\cite{wu2025danet} combines dynamic alignment with CL to predict bulk gene expression from histology. Furthermore, HAGE~\cite{dang2025hage} leverages embeddings from the UNI~\cite{chen2024towards} and introduces hierarchical clustering to align gene-guided image patches with expression embeddings at both local and global scales using CL. Collectively, these approaches indicate that explicitly enforcing correspondence between paired representations can substantially improve mapping fidelity.

Another  challeng in mapping remains robust generalization to unseen tissues, disease contexts, and technical platforms. One proposed strategy is to inject external priors from foundation models. 
FmH2ST~\cite{wang2025fmh2st} extends this idea by modeling both inter-slice heterogeneity and intra-slice complexity with dual feature branches. 
The foundation model branch incorporates images and dual graphs encoding spatial adjacency, using a graph attention network initialized from a pathology foundation model~\cite{hua2024pathoduet}. 
The spot-specific branch extracts local morphological information from H\&E images.
An adaptive attention mechanism 
dynamically
fuses the two branches for prediction. 
The effectiveness of such paradigms naturally depends on the quality and domain coverage of the underlying foundation models. 

Mapping has also been extended beyond transcriptomics to SP, linking morphological patterns to protein-level information.
VirtualMultiplexer~\cite{pati2024accelerating} is an early work as a generative toolkit to synthesize multiplexed IHC images for several antibody markers (AR, NKX3.1, CD44, CD146, p53, and ERG) directly from H\&E images, laying the groundwork for extending to more complex mIF prediction.
ROSIE~\cite{wu2025rosie} predicts mIF protein signals from H\&E images using paired H\&E and CODEX~\cite{van2022multiplex} images from the same specimens, aligned via scale-invariant feature transform and random sample consensus (RANSAC)-based~\cite{fischler1981random} affine alignment. 
While ROSIE demonstrates the feasibility of inferring proteomic patterns from morphology, it highlights practical limitations in cross-site and cross-technology generalization. 
GigaTIME~\cite{Jeya2025gigatime} advances this direction through a multimodal framework that generates virtual mIF images across 21 tumor immune microenvironment markers from H\&E patches, using a NestedUNet-based cross-translator. Trained on paired H\&E–mIF data comprising tens of millions of cells, GigaTIME learns channel-wise activation maps that are stitched into slide-level virtual mIF images, enabling population-level analyses of protein activation patterns and their associations with genomic biomarkers, staging, and survival. 
HEX~\cite{li2026ai} is a recent 
framework that 
generates virtual SP profiles directly from histopathology slides. Leveraging pretrained pathology foundation models, HEX predicts the spatial expression of 40 protein markers,
spanning immune, structural, and functional categories. Extensive validation, together with downstream analyses of prognosis and immune response supported by biological interpretation, highlights its potential for clinical translation to advance precision medicine applications.
Collectively, these methods highlight the progressive expansion of H\&E-to-X mapping from ST to SP, reinforcing histology as a unifying anchor for cross-modal spatial inference.

Mapping methods successfully address the problem of lacking paired modalities to infer molecular profiles from H\&E images, which is promising for future clinical deployments for cost deduction.
However, challenges in generalization capabilities across batches, techniques, and organs still limit inference performance. 

\subsection{Foundation models}
Compared with paired integration and cross-model mapping models trained on 
cohort-specific datasets, 
foundation models aim to learn 
generalizable
representations for diverse downstream tasks 
(e.g.,~\cite{lin2024st,
chen2025visual,
zhang2025histology,
wang2025scgpt,
yang2025past,
tejada2025nicheformer,
blampey2025novae
}).
However, the training of foundation models faces a key challenge: the limited availability of sufficiently large and standardized datasets, particularly paired H\&E and SO data, due to experimental cost and protocol heterogeneity. 
Recent advances have focused on ST, where data volume and consistency are more attainable and have evolved from spot-level toward cellular resolution.

Because ST technologies jointly produce H\&E images and sequencing readouts, 
they naturally lend themselves to the development of foundation models at the spot level.
ST-Align~\cite{lin2024st} is an early image-gene foundation model,
which designs a spot-level and niche-level learning strategy within spatial context.
OmiCLIP~\cite{chen2025visual} leverages CLIP-style architecture~\cite{radford2021learning} with dual image and text encoders, jointly trained on 
2.2 million image-transcriptomic pairs from 32 organs
using a contrastive loss. 
Built on OmiCLIP, the Loki platform 
operationalizes the learned multimodal space for end-to-end ST analysis, including alignment, annotation, and retrieval.
While ST-Align and OmiCLIP emphasize explicit image-gene pairing, 
a complementary strategy is to scale 
pretraining by using H\&E images alone as a universal anchor. 
SpaFoundation~\cite{zhang2025histology} is a visual foundation model trained solely on H\&E images for ST prediction.
Together, ST-Align, OmiCLIP, and SpaFoundation outline a trajectory from  multimodal contrastive alignment to scalable image-only pretraining.

A parallel thread of work aims to learn representations that 
approximate single-cell molecular profiles.
The scGPT-spatial~\cite{wang2025scgpt} extends scGPT~\cite{cui2024scgpt}, a single-cell multi-omics foundation model, to the spatial domain by continual pretraining
on a curated 
ST dataset with 30 million cells or spots with spatial coordinates from 821 individual slides. The key architectural innovation is the introduction of an MoE (Mixture of Experts) decoder to capture modality-specific features.
%
%
PAST~\cite{yang2025past} adopts a CLIP-style dual-encoder contrastive learning architecture~\cite{radford2021learning} and bridges morphology and transcriptomics at single-cell resolution. 
It is trained on 20 million paired H\&E patches and single-cell transcripts from 15 cancer types. 
In addition to typical ST downstream tasks, 
PAST integratively learns morpho-molecular features for multimodal survival prediction, thereby supporting patient-level risk stratification.
%
%
Nicheformer~\cite{tejada2025nicheformer} is a transformer-based ST foundation model pretrained on a collection of single-cell transcriptomics and ST datasets covering over 110 million cells from humans and mice from 73 different tissues and organs. 
Gene expression profiles and metadata of each cell are tokenized using a similar strategy to Geneformer~\cite{theodoris2023transfer} as inputs to the following Nicheformer transformer block to predict masked tokens forming the embedding space. This design yields unified cell embeddings that capture both dissociated single-cell and in situ spatial context and can be transferred across tissues and species. 
Novae~\cite{blampey2025novae} is a self-supervised graph attention network that encodes local environments into spatial representations pretrained on a large ST dataset comprising 30 million cells across 18 tissues. 
It 
focuses on addressing three limitations: (i) methods' reliance on predefined gene panels; (ii) sensitivity to batch effects; and (iii) dependence on external tools for batch-effect correction. Novae deals with the problem of lack of paired datasets by enabling H\&E images as optional inputs.
Together, scGPT-spatial, PAST, Nicheformer, and Novae illustrate a complementary line of foundation models that primarily operate on expression profiles while encoding spatial neighborhood information. 

Beyond ST, recent efforts have extended foundation models to SP. KRONOS~\cite{shaban2025foundation} and Eva~\cite{liu2025modeling} are two examples of this emerging class. KRONOS is pretrained in a self-supervised manner on 47 million SP image patches covering 16 tissue types and 8 fluorescence-based imaging platforms. Built on a ViT backbone, it introduces a third marker-encoding vector to distinguish protein markers across patches, addressing the fact that SP measures far more protein channels than standard RGB images. Eva~\cite{liu2025modeling} is also ViT-based but adopts a masked autoencoder with a hierarchical framework to capture inter-channel and spatial relationships in mIF images. It is pre-trained on paired histopathology-SP data from over 4,000 tissue regions, 64 million cells, and around 200 protein biomarkers. Despite these design differences, both KRONOS and Eva support general-purpose downstream tasks such as image retrieval, tissue classification, and patient stratification, demonstrating the feasibility of foundation models for protein-related spatial analysis.
These advances in SP foundation models motivate the next extension toward SM.

MetaboFM~\cite{ozturk2025metabofm} is built for spatial metabolomics 
on a large corpus of 
mass spectrometry imaging (MSI) data from around 4000 publicly available MSI datasets in the METASPACE~\cite{alexandrov2019metaspace} repository. MetaboFM employs pretrained ViT encoders to learn transferable embeddings of MSI tiles, providing a general representation space 
that can be reused across downstream tasks such as metadata prediction and interactive visual exploration.

Individual foundation models have their own highlights, for example, interpretability, generalization ability, and cross-species or tissue-specific modeling, while there is not a single foundation model that could achieve all highlights, and some foundation models do not fully leverage morphology information from H\&E images. 

In summary, we discuss the evolution trend of integration, mapping, and foundation models from coarse to cellular resolution and from ST to other SO modalities with the anchor of H\&E images. 
Mapping methods to infer molecular information aim to address paired modalities' limitations of integration methods. 
Foundation models are designed to learn robust and comprehensive embeddings from massive available datasets and fine-tune for diverse downstream tasks. 
In reverse, comprehensive representations or feature extractors from foundation models
benefit integration and mapping methods. 
Among the evolving computational methods, 
the role of H\&E images has gradually changed 
from spatial backgrounds providing coordinates to morphological input modalities as representations. 
Together, future directions of SO computational methods are to use available datasets as much as possible and/or design architectures addressing current limitations to improve performance.

\begin{figure}[t]
    \centering
    \includegraphics[width=\textwidth, trim=1.7cm 4cm 1.8cm 1.8cm, clip]{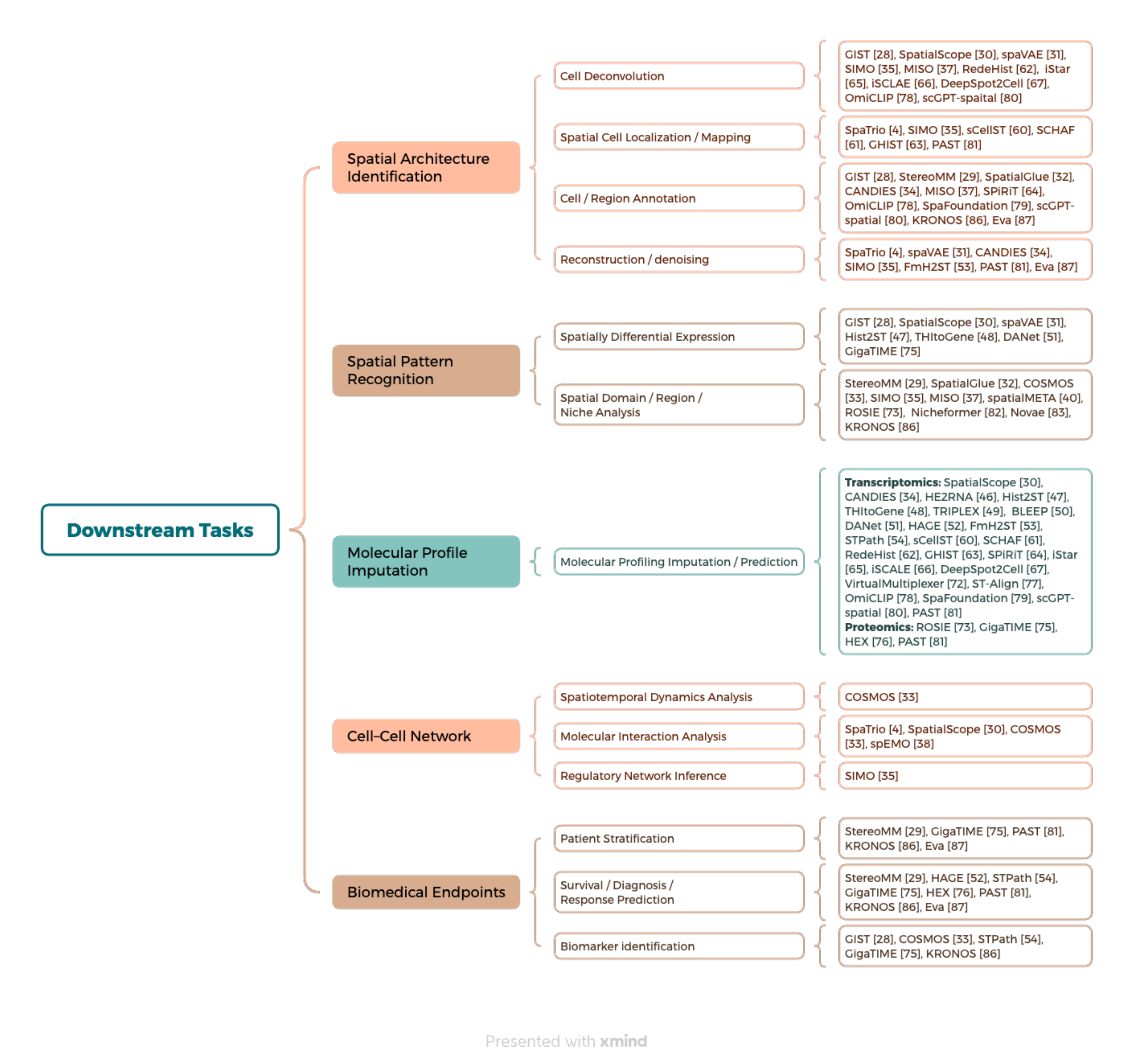}
    \caption{Downstream tasks of computational methods.
We categorize downstream tasks into five major classes, encompassing spatial architecture identification, functional spatial pattern recognition, molecular profile imputation, cell–cell network inference, and biomedical endpoint prediction. 
    }
    \label{fig:tasksF2}
\end{figure}

\section{Downstream tasks and actionable modeling improvements}

\subsection{Downstream tasks for H\&E-centered spatial analysis}
Fig.~\ref{fig:tasksF2} summarizes the main downstream biological and clinical tasks enabled by SO.
We categorize these tasks
into five major classes: spatial architecture identification, functional
spatial pattern recognition, molecular profile imputation, cell-cell network inference,
and biomedical endpoint prediction.

\textbf{Spatial Architecture Identification} constitutes the foundational layer of downstream analysis in SO, aiming to recover the structural organization of tissues from spatially indexed measurements. Typical tasks include cell deconvolution from spot-level data, spatial cell localization, region or layer annotation, tissue reconstruction, and denoising. These analyses establish the spatial scaffold to build higher-level functional and molecular interpretations. 

\textbf{Spatial Pattern Recognition} builds on reconstructed tissue architecture to identify functionally coherent spatial domains or molecular patterns beyond structural organization. This includes detecting spatially differentially expressed genes, delineating functional regions, and characterizing cellular niches shaped by local microenvironments. By linking molecular variation to spatial context, pattern recognition reveals emergent tissue functions that are not observable from non-spatial data.

\textbf{Molecular Profile Imputation} focuses on inferring missing or unmeasured molecular signals, such as gene or protein expression, from available modalities. This task is particularly critical given the sparsity, noise, and cost constraints of SO technologies. From a modeling perspective, H\&E-to-X mapping methods can be viewed as specialized imputation frameworks that leverage morphology to recover high-resolution molecular profiles.

\textbf{Cell-Cell Network Inference} becomes feasible as spatial resolution approaches the cellular or subcellular level. These analyses aim to reconstruct cell-cell communication networks, molecular interaction pathways, and regulatory circuits directly in situ, often extending to spatiotemporal dynamics~\cite{zhou2025cooperative}. By embedding cellular interactions within their native spatial context, such models provide mechanistic insights into tissue organization, development, and disease progression. 

\textbf{Biomedical Endpoint Prediction} represents the translational capability of SO analysis, connecting spatial molecular patterns to clinical and biological outcomes. Applications range from patient stratification and survival analysis to biomarker discovery and treatment response prediction. Anchored by H\&E images as the clinical gold standard, these tasks highlight the growing potential of SO to inform decision-making in precision medicine.

Beyond methods, comprehensive platforms built on spatial foundation models have emerged to support diverse downstream analyses. For example, the Loki platform is built on the backbone of OmiCLIP and allowed multimodal downstream analysis for diverse annotations, cross-modality imputations, etc. Similarly for SP, the KRONOS provides embeddings that could be used for classification, clustering identification, and prediction for phenotyping and biomedical endpoints, including biomarker discovery and patient stratification.

\subsection{Common Failures and Actionable Directions}

While the downstream tasks discussed above demonstrate the vast potential of SO, their success is often contingent upon the quality and scale of the underlying data. Current computational frameworks have introduced several architectural innovations to address inherent technical limitations, such as coarse resolution and batch effects. In this section, we categorize 
``Common Failures'' with actionable directions as follows: (i) resolution-related failures, (ii) generalization-related failures, and (iii) interpretation- and evaluation-related failures.

\textbf{Resolution-related failures} largely arise from a fundamental mismatch between 
the measurement granularity of  
SO technologies and 
the cellular (or subcellular) resolution required by downstream biological questions. 
In practice, a substantial fraction of datasets remain spot-level or only ``pseudo'' cell-level, capturing mixed signals from multiple neighboring cells and interstitial material, thereby reducing the fidelity of downstream analyses. 
SR for ST~\cite{wan2023integrating, wang2025fmh2st,nonchev2025deepspot2cell} leverages abundant scRNA-seq references and pathology foundation model embeddings to infer virtual single-cell expression from 
H\&E images.
However, analogous non-transcriptomic SR remains limited despite H\&E-to-SP feasibility shown by ROSIE~\cite{wu2025rosie}, GigaTIME~\cite{Jeya2025gigatime}, HEX~\cite{li2026ai}, and Eva~\cite{liu2025modeling}, largely due to scarce paired multimodal datasets. 
Furthermore, quantitative agreement with ground truth for current cellular-scale reconstructions can remain modest (e.g., GHIST reports Pearson correlation coefficients below 0.2 between predicted and measured gene expression).
These failures are driven by compounding technical and algorithmic factors. 
H\&E image quality is intrinsically compromised due to technique workflows when H\&E staining is performed after iterative hybridization or imaging, leading to reduced contrast, tissue distortion, and, consequently, ambiguous cell boundaries.
%
Thus, accurate cell segmentation becomes particularly challenging when applied to sparse or low-quality H\&E images for weak boundaries and complex tissue structures (e.g., axons or irregular cell morphologies). 
Since these inaccuracies are propagated and amplified during representation learning, especially at the cellular scale, developing robust segmentation frameworks capable of handling vague margins and irregular shapes is a critical architectural priority.

\textbf{Generalization-related failures} are the inability of SO computational methods to maintain reliable performance on unseen samples beyond the training distribution. 
We categorize them as the following levels:
\begin{itemize}[noitemsep,topsep=5pt]

\item \textit{Scale Generalization} (intra-slide and inter-slide): Standard H\&E WSIs (approx. 25 mm $\times$ 55 mm) significantly exceed the capture area of current SO techniques. 
Only selected FOVs 
have spatial profiles, 
leading to sampling bias and missing key biological regions.
Therefore, models should be capable of extrapolating molecular profiles from restricted FOVs to WSIs. For instance, iSCALE~\cite{schroeder2025scaling} can expand gene expression coverage from FOVs to the entire tissue section, though its performance remains sensitive to 
the
FOV selection. Beyond intra-slide expansion, methods should also consider identifying inter-slide domains to discover common biological features and spatial biomarkers across samples.

\item \textit{Batch and Technical Generalization}:
Batch effects (i.e., domain shifts) introduce non-biological variation across samples in tissue preparation, experimental batches, and platform-specific biases, and they can confound true biological heterogeneity.
%
Current frameworks address these challenges through several strategies~\cite{marconato2025spatialdata, tian2024dependency,liu2025cross,blampey2025novae}. 
First, some methods explicitly model noise within the architecture to decouple biological signals from technical artifacts~\cite{tian2024dependency,liu2025cross}. 
Second, self-supervised alignment strategies have proven effective. For example, Noeva~\cite{blampey2025novae} utilizes a self-supervised graph attention network 
to align cells with similar biological features in the embedding space
to address the 
batch effects 
problem. 
Third, batch effects can be minimized by conditioning the model on explicit metadata, such as tissue origin, technology platform, and sequencing protocols
~\cite{park2024cellama, wang2025scgpt, tejada2025nicheformer}. 
scGPT-spatial~\cite{wang2025scgpt} also avoids explicitly encoding spatial coordinates to ensure cross-slide generalizability. 
Together, systematic integration of domain-aware objectives is essential to ensure that learned embeddings remain biologically meaningful.

\item \textit{Organ and Species Generalization}: 
Models often struggle in unseen organ or disease contexts 
due to highly tissue- and disease-specific expression programs and morphology-molecular associations.
For example, STPath~\cite{huang2025stpath} shows reduced accuracy in renal cell carcinoma gene expression prediction.  
This limitation may be mitigated 
by leveraging 
pathology 
and 
single-cell foundation models pretrained on diverse, multi-site, and multimodal datasets as frozen feature extractors 
within the 
architecture
~\cite{lin2024st, wang2025fmh2st, nonchev2025deepspot2cell}.
Beyond organ transfer, 
emerging work approaches species generalization by 
learning shared biological embeddings that capture conserved mechanisms across species (e.g., joint human-mouse brain representations~\cite{zhang2025brainbeacon}), 
enabled by large-scale cross-species references such as the Brain Cell Census~\cite{BrainCellCensus}.
Scaling these resources and improving cross-domain alignment will support robust transfer and mechanistic interpretation 
across tissues and species.

\end{itemize}

\textbf{Interpretation- and evaluation-related failures} arise when
models produce accurate-looking predictions but provide limited biologically meaningful rationales and are assessed under benchmarks that do not faithfully reflect the target biological questions. 
As 
moving
toward clinical deployment, models must transcend black-box outputs by supporting mechanistic or histological explanations. 
For example, SPiRiT~\cite{zhao2024inferring} integrates the TAVAC metric into a ViT architecture to guide attention toward informative regions and report the areas and attention details used to predict marker gene expression.
Robust clinical adoption also requires rigorous multi-site validation and generalization following standardized guidelines for analytical reporting~\cite{lee2025spatial}. 
Current studies often use restricted datasets with inconsistent experimental settings, and the chosen metrics poorly match the biological tasks, leading to potentially inflated performance estimates. 
Addressing these limitations calls for unified, large-scale benchmarks spanning diverse cohorts and platforms, together with biologically informed evaluation metrics and standardized reporting practices, to ensure that methodological advances translate into actionable biological insights and clinically deployable tools.

In summary, no single framework yet resolves all failure dimensions simultaneously. Importantly, these limitations do not imply that H\&E-centered modeling is fundamentally infeasible. Instead, they largely reflect practical constraints, including insufficient (and weakly paired) data, restrictive modeling assumptions, and evaluation protocols misaligned with underlying biological questions. 
These failure modes motivate actionable directions for method development while set the stage for distinguishing correctable gaps from the more intrinsic limitations discussed next.


\section{Discussion and conclusion}

We have discussed the computational trajectory of SO, evolving from multimodal integration to cross-modal mapping, and the emergence of foundation models. 
This 
modeling
maturation coincides with a shift in downstream applications 
from 
biologically
mechanistic explorations 
toward clinically oriented endpoints. Central to this evolution is the systematic utilization of H\&E images as a morphological anchor to unify heterogeneous modalities and push toward 
cellular
or subcellular resolution. 
While architectural innovations have addressed many computational hurdles, several fundamental bottlenecks rooted in biological complexity and data acquisition constraints remain unresolved, as outlined below:

\begin{itemize}[noitemsep,topsep=5pt]
\item \textbf{Representation Gap.} 
A fundamental bottleneck in current SO is the 
``representation gap'' 
between measured data and biological reality.
Most methods implicitly assume that selected FOVs are representative, reflecting underlying pathological and physiological states. 
However, typical tumor volumes and 
tissue architectures substantially exceed the sampled regions and the resolution of 
H\&E and SO measurements, 
such that only a fraction of spatial molecular and morphological heterogeneity is captured.
Although SO preserves spatial context relative to bulk or single-cell sequencing, 
key expression patterns or rare biomarkers may remain undetected due to sparse sampling and resolution limits~\cite{schroeder2025scaling}.
%
Addressing this gap will require multi-region, multi-scale modeling and strategic region-of-interest selection to improve coverage of critical biological signals~\cite{yuan2025smart}.

\item \textbf{Dimensionality Gap.} 
Despite the nomenclature of 
``spatial'' omics, 
most existing data remain 
two-dimensional (2D) snapshots of three-dimensional (3D) biological systems. 
Sample preparation and sequencing constraints typically limit measurements along orthogonal axes, 
where the distance between adjacent sections often exceeds single-cell dimensions. 
Together with the fact that only several selected FOVs per section are profiled, 
these limitations
complicates the reconstruction of coherent 3D tissue volumes, 
potentially obscuring long-range cellular interactions and architectural continuity~\cite{almagro2025ai}. 
Bridging this gap will require methodological advances to 
align adjacent sections and interpolate ``vertical'' biological information.
Emerging frameworks such as STAIG~\cite{yang2025staig} and VORTEX~\cite{almagro2025ai} have begun to address these challenges. 
As sequencing costs decrease and alignment algorithms improve, this field may move toward
3D and even 4D longitudinal (space and time) representations, enabling characterization of dynamic processes such as tumor evolution and treatment response in their native volumetric context.

\item \textbf{Technology Gap.} 
A critical challenge 
in multimodal learning 
is the limited ability to capture multiple modalities from the same tissue section. 
Although some methods attempt to integrate data across sections, 
a ``gold standard'' in which different modalities are perfectly aligned is lacking.
%
This limitation could be resolved by the development of consecutive profiling or same-section multimodal profiling techniques.
%
In addition, 
H\&E 
image  
quality remains an often overlooked challenge. 
As detailed in Section~\ref{sec:tech}, 
H\&E images from SO experiments are frequently suboptimal due to chemical damage 
incurred during cyclic staining and bleaching, 
resulting in lower quality than diagnostic histology.
Such degradation poses a significant hurdle for computational architectures and clinical deployment.
%
Furthermore, SP technologies such as mIF imaging are limited by the number of antibody markers that can be profiled per section, restricting protein coverage relative to the transcriptome. 
Imaging-based ST methods likewise profile  far fewer genes than whole-transcriptome sequncing-based approaches.
\end{itemize}

In summary, this paper provides a comprehensive and structured overview of the rapid computational evolution in SO, particularly focusing on multimodal analysis anchored by histopathology at single-cell resolution. 
By delineating current architectural advancements from persistent systemic bottlenecks, we provide a framework to understand the field's progress and considerations in method development: what is currently achievable through model design and what remains constrained by fundamental technological limits. 
Together, this paper serves as a 
roadmap of SO computational methods for both computational biologists developing next-generation models and translational researchers seeking to deploy these methods in clinical settings.

\bmhead{Supplementary information}
\textbf{}




\textbf{Additional file 1: Table S1.} Summary of the main datasets utilized by the reviewed computational methods.
The collection is organized by tissue cohort, data modality, and the underlying 
technology used to generate the data.

\textbf{Additional file 1: Table S2.} Summary of the reviewed computational methods. This table provides a comprehensive overview of the Integration, Mapping, and Foundation Models discussed in this survey, detailing their publication year, venue, applicable data modalities, resolution level (spot vs. single-cell), and model architecture.

\newpage




\section*{Declarations}


\subsection*{Funding}

G.W. is supported by the National Cancer Institute of the National Institutes of Health under Award Number K99CA286966.

\subsection*{Competing interests} 

Not applicable.

\subsection*{Ethics approval and consent to participate}

Not applicable.

\subsection*{Consent for publication}

Not applicable.

\subsection*{Availability of data and materials}

Not applicable.

\subsection*{Author contributions}

N.H., X.Y., C.Z., and G.W.: Conceptualization, Design, Methodology, Investigation. 
N.H., X.Y., B.S., D.L., C.Z., and G.W.: Original Draft.
C.Z. and G.W.: Supervision.
G.W.: Funding, Resources, and Project Administration.
All authors reviewed and approved the final manuscript.








\bibliography{sn-bibliography}









\end{document}